# Dynamic origin of the morphotropic phase boundary - Soft modes and phase instability in $0.68Pb(Mg_{1/3}Nb_{2/3}O_3)$-$0.32PbTiO_3$

*Hu Cao[1], Chris Stock[2], Guangyong Xu[3], Peter M. Gehring[4],*

*Jiefang Li[1], and Dwight Viehland[1]*

[1]Department of Materials Science and Engineering, Virginia Tech, Blacksburg, VA 24061, USA

[2]ISIS Facility, Rutherford Appleton Labs, Chilton Didcot, Oxon, United Kingdom OX11 0QX

[3]Condensed Matter Physics and Materials Science Department, Brookhaven National Laboratory, Upton, NY, 11973

[4]NIST Center for Neutron Research, National Institute of Standards and Technology, Gaithersburg, MD 20899-6100, USA

We report neutron inelastic scattering and high-resolution x-ray diffraction measurements on single crystal $0.68Pb(Mg_{1/3}Nb_{2/3}O_3)$-$0.32PbTiO_3$ (PMN-0.32PT), a relaxor ferroelectric material that lies within the compositional range of the morphotropic phase boundary (MPB) where the piezoelectric properties of PMN-$x$PT compounds are close to maximal. Data were obtained between 100 K and 600 K under zero and non-zero electric field applied along the cubic [001] direction. The lowest energy, zone-center, transverse optic phonon is strongly damped and softens slowly at high temperature; however the square of the soft mode energy begins to increase linearly with temperature as in a conventional ferroelectric, which we term the soft mode "recovery," upon cooling into the tetragonal phase at $T_C$. Our data show that the soft mode in PMN-0.32PT behaves almost identically to that in pure PMN, exhibiting the same temperature dependence and recovery temperature even though PMN exhibits no well-defined structural transition (no $T_C$). The temperature dependence of the soft mode in PMN-0.32PT is also similar to that in PMN-0.60PT; however in PMN-0.60PT the recovery temperature equals $T_C$. These results suggest that the temperature dependence and the energy scale of the soft mode dynamics in PMN-$x$PT are independent of concentration on the Ti-poor side of the MPB, but scale with $T_C$ for Ti-rich compositions. Thus the MPB may be defined in lattice dynamical terms as the concentration where $T_C$ first matches the recovery temperature of the soft mode. High-resolution x-ray studies show that the cubic-to-ferroelectric phase boundary shifts to higher temperatures by an abnormal amount within the MPB region in the presence of an electric field. This suggests that an unusual instability exists within the apparently cubic phase at the MPB.

I. Introduction

Solid solutions of $(1-x)$Pb(Mg$_{1/3}$Nb$_{2/3}$O$_3$)-$x$PbTiO$_3$ (PMN-$x$PT) and $(1-x)$Pb(Zn$_{1/3}$Nb$_{2/3}$O$_3$)-$x$PbTiO$_3$ (PZN-$x$PT) belong to the class of materials known as relaxor ferroelectrics and have attracted intense interest from both scientific and engineering communities over the past decade because of the enormous potential they possess for use in high performance, piezoelectric actuator and transducer applications [1]. Historically, the excellent electromechanical properties of the closely related Pb(Zr$_{1-x}$Ti$_x$)O$_3$ (PZT) ceramics were attributed to the steep morphotropic phase boundary (MPB) that separates Ti-poor rhombohedral (R) and Ti-rich tetragonal (T) ferroelectric phases as a function of the Ti content $x$ [2]. In 1997 Park and Shrout [3] conjectured that the ultrahigh strains they achieved in <001> oriented crystals of PMN-$x$PT and PZN-$x$PT, which exhibit similar MPBs, resulted from an R→T transition induced by an applied electric field E.  However various intermediate monoclinic phases (M$_C$, M$_A$, and M$_B$), which provide a natural bridge connecting the R and T phases, have subsequently been reported near the MPB in both PZT ceramics [4-6] and in single crystals of PZN-$x$PT [7-11] and PMN-$x$PT [11-16]. It has since been argued that monoclinic symmetry may be an important property underlying the exceptional piezoelectric response of PMN-$x$PT and PZN-$x$PT as it allows the polarization vector to rotate freely within a plane [17] rather than be restricted to a particular crystallographic axis as is the case for the higher symmetry R, T, and orthorhombic (O) phases.

Neutron scattering studies have advanced our understanding of PMN-based relaxor ferroelectrics significantly [18-24], however to date comparatively few lattice dynamical measurements have been made on PMN-$x$PT crystals with compositions lying within the MPB. Considering the precipitous drop in the piezoelectric coefficients that occurs when crossing from the Ti-poor to the Ti-rich side of the MPB, such measurements are of obvious importance to efforts to determine the microscopic mechanism responsible for the anomalously high electromechanical response of these materials. In conventional ferroelectrics such as PbTiO$_3$ (PT), neutron inelastic scattering was used to show that the condensation or softening of a zone-center, transverse optic (TO) phonon mode drives the C→T ferroelectric transition [25], while the square of the soft mode energy $(\hbar\omega_0)^2$ varies linearly with temperature above and below the Curie point ($T_C$ = 763 K) according to the Lyddane-Sachs-Teller (LST) relationship [26]. Neutron studies of the relaxors PMN [19,20], PZN [27], and PZN-0.08T [28], however, revealed notable differences with $PbTiO_3$ in that the corresponding TO mode softens slowly and becomes heavily damped at high temperature near $T_d$, the well-known Burns temperature where static, polar nanoregions (PNRs) first appear [29]. The soft mode remains damped down to ~210 K, the temperature below which a remnant polarization can be sustained in a field-cooled state [30], but exhibits an undamped lineshape at lower temperatures. In addition, neutron inelastic scattering studies of PZN [27] and PMN [20] have shown that this unusual damping extends to reduced wave vectors $q \leq 0.2$ Å$^{-1}$, a dynamical feature that is now known as the 'waterfall' effect in which long-wavelength TO modes exhibit a strongly $q$-dependent damping. Subsequent neutron studies of PMN-0.20PT [22] and PMN-0.60PT [24] reported similarly damped soft modes between $T_d$ and $T_C$ for wave vectors near the zone center.

Here, we study the lattice dynamics of field cooled (FC) PMN-0.32PT (for E=1kV/cm) using neutron elastic and inelastic scattering techniques, and compare our results to those previously reported for PMN [19-21, 23] and PMN-0.60PT [24]. For pure PMN the soft mode recovers near 400 K, a temperature well above 210 K where a FC (field-cooling) structural transition is observed. By contrast, in heavily doped PMN (PMN-0.60PT) these two temperature scales coincide. In this work we investigate the behavior of the soft, zone center optic phonon in PMN-0.32PT and show that the MPB defines a concentration where the structural transition and the mode recovery occur at the same temperature.

Our paper differs from a recent neutron investigation of ZFC (zero-field-cooling) PMN-0.32PT (Ref [31]), which focused on the low-energy, quasielastic scattering associated with polar fluctuations and did not track the soft mode through the critical temperature or give a detailed comparison of the soft mode behavior to that in PMN and PMN-0.60PT. In this paper we show that the zone center, transverse optic phonon in PMN-0.32PT is soft over a broad temperature range within both the C and T phases. Despite the fact that the structural properties change dramatically with temperature, we find that the lattice dynamics are nearly identical to those in the prototypical relaxor PMN. The dynamics are also similar to those in PMN-0.60PT, but only after they have been scaled by the critical temperature $T_C$. These results thus suggest the remarkable result that the lattice dynamical properties of PMN-$x$PT are universal for concentrations below the MPB.

## II. Experimental Details

Three (001)/(110)/(1-10) bar-shaped (3×3×9 mm$^3$) PMN-0.32PT crystals were purchased from Academic Sinica (Shanghai, China) [32]. The specimens were cut along pseudocubic crystallographic {100} faces to within a tolerance of ±0.5°. Gold electrodes were deposited on one pair of opposite (001) faces by sputtering. The temperature dependent dielectric constant was measured for each crystal using an HP 4284A LCR meter, from which it was insured that $T_{max}$=412±3°C. High resolution x-ray diffraction (XRD) measurements were performed on one of the crystals using a Philips MPD system equipped with a double-bounce, hybrid monochromator, an open 3-circle Eulerian cradle, and a domed hot-stage. A Ge (220)-cut crystal was used as an analyzer, which had an angular resolution of 0.0068˚. The x-ray wavelength was that of Cu$_{K\alpha 1}$=1.5406 Å, and the x-ray generator was operated at 45 kV and 40 mA. Each measurement cycle was begun by heating to 550 K to depole the crystal; measurements were then taken on cooling.

Neutron scattering experiments were performed at the NIST Center for Neutron Research using the BT7 triple-axis spectrometer, which is equipped with a variable vertical focus monochromator and a flat analyzer. Three PMN-0.32PT crystals were co-aligned to within <1° to give an effective mass of 1.9 g. The room temperature lattice constant $a$ = 4.004 Å; thus one reciprocal lattice unit (rlu) equals 1.569 Å$^{-1}$. The (002) reflection of highly-oriented, pyrolytic graphite (HOPG) crystals was used to monochromate the incident neutron energy $E_i$ and to analyze the final neutron energy $E_f$.

The data were taken using a fixed final neutron energy of $E_f$ = 14.7meV (λ = 2.36Å) and horizontal beam collimations of open-50'-S-40'-120' ("S" = sample) between reactor and detector. An HOPG transmission filter was placed in the scattered beam after the analyzer to eliminate high-order neutron wavelengths. The co-aligned crystals were mounted onto an aluminum holder and oriented with the cubic [001] and [110] axes in the horizontal plane to achieve an [HHL] scattering zone. The sample was then placed inside a vacuum furnace capable of reaching 600 K, and an electric field was applied along [001] while cooling below 450 K. The application of an electric field pins one of the crystallographic directions on cooling through the various structural transitions and therefore reduces the effective mosaic at low temperatures introduced through the domain formation. Since low-energy transverse phonons are sensitive to the presence of domains, this procedure allowed us to measure the transverse acoustic and optic phonons close to the zone center.

Constant-**Q** scans were performed in the (220) Brillouin zone along the [001] direction by holding the momentum transfer **Q** = **k**$_i$-**k**$_f$ ($k$ = 2π/λ) fixed while varying the energy transfer $\Delta E = \hbar\omega = E_i$-$E_f$. Using these scans the dispersion of both the transverse acoustic (TA) and the lowest-energy transverse optic (TO) phonon branches were determined at various temperatures. The data were corrected for higher-order contamination as described elsewhere [33]. To investigate the effect of mode coupling, we carried out a detailed lineshape analysis [21, 23]; the linewidth, amplitude, and phonon frequency were obtained by fitting the data to a Lorentzian lineshape characteristic of a damped harmonic oscillator convolved with the instrumental resolution function. A Gaussian lineshape was used to describe the scattered intensity at $\Delta E$ = 0, and a constant was used to describe the overall background.

### III.  Neutron inelastic scattering results for PMN-0.32PT

To extract physically meaningful parameters such as phonon frequency and linewidth, a model neutron scattering cross section must be convolved with the instrumental resolution function and fit to the data. We fit our data to a model using two damped harmonic oscillators to describe the TO and TA modes as done in previous studies and outlined in detail elsewhere [21, 23]. The results of these fits are discussed in the following sections.

Figure 1 shows constant-$Q$ scans at (2, 2, 0.15) and (2, 2, 0.2) measured at 450 K, 390 K, and 100 K, temperatures that correspond to the C, T, and $M_C$ phases, respectively. In the C and T phases we found no propagating soft, transverse optic (TO) modes at the zone center ($q = 0$). Even scans measured at small but non-zero wave vector revealed no clear indication of a TO mode in either phase because of the heavily damped nature of the scattering cross section. Evidence of a TO mode only became apparent at wave vectors $q > 0.1$ rlu at which point a broad TO mode was observed. On cooling through the C→T phase transition at $T_C$, a weak recovery of the optic mode was seen; it is only at low temperatures that an underdamped optic mode was observed. These measurements are in accord with those on pure PMN and PMN-60PT, which are summarized in Fig. 2. For example, at 100 K a well-defined TO mode at $q = 0.1$ rlu is present in each compound; however at higher temperatures no well-defined propagating mode is visible near the zone center. On the other hand, well-defined TA modes are observed over the entire temperature range studied, while the TA frequency changes relatively little with temperature.

The constant-$Q$ scans at $q = 0.10, 0.15$, and $0.20$ rlu demonstrate the presence of strong TO mode damping near the zone-center in both the T and C phases. While it is possible that the degree of damping of the zone center soft phonon is larger than that in PMN [37-39] and PMN-60PT [20], the temperature dependence and energy scale of the zone center soft mode are nearly identical to those of the soft mode in PMN and scale with the critical temperature $T_C$ for PMN-60PT. Together these results suggest a common dynamic response throughout the PMN-$x$PT phase diagram, which is remarkable considering the very different structural properties and phase transition sequences that take place with increasing PT concentration.

### A. Observation and study of the zone-center, soft mode

Figure 2(b) shows the square of the zone-center, soft mode energy $(\hbar\omega_0)^2$ as a function of temperature for PMN-0.32PT. These data are compared to previously published results for PMN and PMN-0.60PT [24], which are shown in Figs. 2(a) and (c) respectively. The open circles represent fitted parameters derived from constant-$Q$ scans measured at the zone center, whereas the solid circles represent values obtained via extrapolation from constant-$Q$ scans measured at nonzero $q$ using the formula $\omega^2=\omega_0^2+\alpha q^2$, where $\omega_0$ is the zone-center frequency and $\alpha$ is a temperature independent constant. This second method, which has been used to analyze the soft mode behavior of PbTiO$_3$ [29], is required in the waterfall region near the zone center where the optic mode becomes extremely broad in energy and is difficult to discern in a constant-$Q$ scan.

It can be seen from Fig. 2 that the zone-center, soft TO mode for the three PMN-$x$PT crystals does not completely soften to zero energy near the critical temperature as it would for a second-order phase transition (the transition is first-order in pure PbTiO$_3$, so the soft mode energy does not reach zero in that case either). Rather, the zone-center, soft mode for each of these three crystals is heavily damped over a large temperature range near and above T$_C$. For PMN-0.60PT the minimum in the zone-center, soft mode energy corresponds to the C→T transition, which occurs at T$_C$ = 550 K as determined by XRD and dielectric constant measurements [24]; below T$_C$ the soft mode recovers as expected for a conventional ferroelectric. For PMN, however, the minimum in the zone center, soft mode energy does *not* correspond to an observable structural phase transition, as was also confirmed by XRD and dielectric measurements [34]. Nevertheless the soft mode in PMN also recovers at low temperatures like a ferroelectric. In between these two extremes we find that for PMN-0.32PT the zone-center, soft mode minimum energy occurs near the C→T transition at 410 K. We further note that we observe no obvious dynamic signature of the transition to the monoclinic phase at lower temperature; instead the temperature dependence of the soft mode is, remarkably, nearly identical at all temperatures below T$_C$ to that in the relaxor PMN, which exhibits short-range polar correlations, and to that in the ferroelectric PMN-60PT, which exhibits long-range polar correlations. By contrast, the elastic diffuse scattering varies significantly across the PMN-$x$PT phase diagram and actually disappears for PT compositions beyond the MPB boundary, where the relaxor character of the dielectric response is replaced by a well-defined peak characteristic of a conventional ferroelectric phase [35].

A comparison of the temperature dependence of the zone-center, soft mode energy in PMN, PMN-0.32PT, and PMN-0.60PT reveals three interesting results. First, the overall behavior of $(\hbar\omega_0)^2$ versus temperature is very similar with the low temperature slope decreasing slightly with increasing PT content. Second, the temperature at which the soft mode begins to recover is almost identical for PMN and PMN-0.32PT, even though PMN-0.32PT exhibits two well-defined structural phase transitions and no structural transition has been unambiguously identified in the bulk of PMN (in either zero or non-zero applied electric field) using neutron or x-ray diffraction techniques [36-40]. Third, the soft mode recovery and the C->T phase transition occur at about the same temperature in both PMN-0.32PT and PMN-0.60PT, with the transition temperature $T_C$ being higher in PMN-0.60PT. Together these results suggest that the MPB defines a PT concentration where the structural transition temperature $T_C$ matches the temperature at which the soft mode recovers. This conclusion will be discussed in more detail later.

### B. TO and TA phonon dispersions

The dispersions of the TO and TA modes for PMN-0.32PT were measured in the (220) zone at 600K (E = 0kV/cm) and 300K ($E_{[001]}$= 1kV/cm) and are shown in Figure 3(a). At large $q > 0.2$ rlu, the TO branch changes relatively little with temperature. For $q \leq 0.2$ rlu, however, the TO branch softens significantly at 600K, which is in the C phase. The dashed line represents the fact that at small $q < 0.15$ rlu well-defined TO modes could not be resolved in constant-$Q$ scans because they were too heavily damped by the waterfall effect. Strong phonon softening and damping persist on cooling from 600K into the T phase. At 300K, the softening of the TO mode is much less pronounced and a phonon peak can be determined to lower values of $q$; this indicates that the TO phonon lifetime increases on entering the T phase. A direct estimate of the zone-center, soft mode energy at 600K is effectively impossible because of the large damping effects. To overcome this limitation, we exploited the fact that the TO dispersion is quadratic in $q$ for small wave vectors, as was done in PbTiO$_3$ [25]. Neglecting the data points at larger values of $q$ = 0.28 and 0.30 rlu, we then performed a linear extrapolation by plotting $(\hbar\omega_0)^2$ vs. $q^2$. The results of the extrapolation are presented in Fig. 3(b), which gives results similar in both absolute energy and temperature dependence to those in PMN [19, 27].

By contrast, the TA dispersion measured is essentially temperature independent and similar to those previously reported for other PMN-$x$PT compositions [20, 22, 25]. Some changes do occur in the TA phonon parameters with temperature, and these are discussed below. The lack of temperature dependence of the TA phonon branch, despite the significant softening of the TO mode over a broad temperature range, is actually a key observation in light of recent models of the relaxor phase transition. When compared to other compounds where strong TA-TO coupling is present [41-43], this finding indicates that there is no or very weak coupling between the TA and TO modes. While this conclusion contradicts those in Refs [31], we note that these other studies addressed neither the temperature dependence nor the lineshape of the TO phonons, both of which have been consistently and well described in pure PMN using a purely harmonic model with two uncoupled damped harmonic oscillators [23].

**C. Temperature dependence of the phonon energy and damping at non-zero $q$**

Here we discuss the temperature dependence of the TA and TO phonon energies and linewidths away from the zone center measured at (2, 2, 0.15) for PMN-0.32PT under $E_{[001]}$=1 kV/cm. Fig. 4(a) shows that the energy of the TO mode increases on cooling below T$_C$, the rate of increase changing only slightly in the temperature range of 300-400 K, and then increases faster on cooling below 300 K. Although within the experimental error, the small change in slope observed between 300 K and 400 K coincides with the structural phase sequence C→T→M$_C$ on cooling illustrated in Fig. 6. As suggested by the data at 300 K and 600 K in Fig. 3(b), the energy of the TA mode is effectively temperature independent from 100 K to 600 K, except near 350 K, in the vicinity of the T→M$_C$ transition, where there is a hint of a shallow minimum that is also well within our

"*Dynamic origin of the morphotropic phase boundary…*" by Hu Cao *et al.*

experimental uncertainties.

The temperature dependence of the linewidth ($\Gamma$) of the TA and TO modes for $q = 0.15$ rlu is given in Fig. 4(b). A large TO mode linewidth ($\Gamma_{TO}$) is seen at 600 K, which decreases linearly by ~50% on cooling to 400 K. This high temperature broadening indicates a surprisingly strong damping (short lifetime) of the TO mode far into the cubic phase. On cooling into the T phase near $T_C$, $\Gamma_{TO}$ becomes relatively independent of temperature. On cooling further into the $M_C$ phase, $\Gamma_{TO}$ decreases substantially with temperature down to 100 K. Therefore, the TO mode linewidth appears to respond to the presence of the two structural transitions in spite of the fact that the zone-center, soft mode energy extrapolates smoothly through the same temperature range. The linewidth of the TA mode ($\Gamma_{TA}$) is much narrower than that for the TO mode over the entire temperature range investigated, and is almost temperature independent. When we compare the linewidths of the TO and TA modes to those measured in pure PMN, we find that the linewidths in PMN-0.32PT are larger, indicating a shorter lifetime and possibly more phase instability, which goes hand-in-hand with the structural result discussed below.

## IV. Dielectric and XRD measurements

Prior dielectric and XRD measurements of PMN-$x$PT over a wide range of compositions have shown that $T_{max}$ (dielectric maximum) is greater than $T_C$ for small $x$, approaches $T_C$ with increasing x, and equals $T_C$ for $x > 0.37$. These results are summarized in Fig. 6. Correspondingly, when $T_{max} > T_C$ strong relaxor characteristics are found in the dielectric response, which becomes increasingly less apparent as $T_{max} \rightarrow T_C$, and which disappear when $T_{max} = T_C$ resulting in normal ferroelectric behavior. In the vicinity of the MPB, weak relaxor characteristics are found in the ZFC state, but not in the FC state [44].

We performed dielectric constant and lattice parameter studies in the FC state to compare with our neutron inelastic scattering results, the data for which are shown in Fig. 5. The temperature dependent dielectric constant ($1/\varepsilon_r$) for $E_{[001]}=1$ kV/cm clearly shows two phase transitions on cooling: one near 410 K and a second at 320 K, as can be seen in Fig.5 (a). High resolution XRD studies determined this transformational sequence to be C→T→$M_C$ as shown in Fig.5 (b), in agreement with the diagram given in Fig. 6. The dielectric constant peak is sharp and exhibits no noticeable frequency dispersion about $T_{max}$ for $E_{[001]}$ = 1 kV/cm, as reported here, or in the ZFC state as previously shown [15]. In this regard, PMN-0.32PT exhibits phase transformational characteristics similar to those of a normal ferroelectric, rather than those of a relaxor. However, linear Curie-Weiss behavior is not observed for $1/\varepsilon_r$ for temperatures above the peak in the dielectric constant; this indicates the presence of a local polarization to temperatures much higher than $T_{max}$, and presumably up to $T_{Burns}$ (620 K).

We next measured the field dependence of the *c*-axis spacing in the vicinity of the C→T boundary for PMN-0.32PT. We found no significant difference in *c*-axis spacing between E = 0 and $E_{[001]}$ = 0.5kV/cm, although the C→T boundary increases by about 5 K. However, at higher applied fields $E_{[001]}$ > 2.5 kV/cm, we found that this boundary shifts rapidly to higher temperatures by about 75 K. Since the *c*-axis spacing increases by application of $E_{[001]}$ above 425 K, we can assume that the crystal does not remain cubic in the temperature range of 425-500 K; instead, the data indicate a lower symmetry structure. The shift of the C→T boundary by $E_{[001]}$ determined by XRD is not accompanied by a corresponding shift in the temperature of the dielectric maximum. Step-like changes in the *c*-axis spacing occur near 425 K and 500 K. For $E_{[001]}$ = 5 kV/cm (data not shown), it was not possible to determine a distinct C→T phase boundary because the *c*-axis spacing changes smoothly and continuously on cooling. We note that the temperature region in which the C→T boundary is significantly shifted by an applied electric field corresponds to that in which heavily damped phonons are observed. Crystals with higher and lower PT contents do not exhibit such a dramatic shift in the C→T boundary (summarized in Fig. 6).

## V.   Discussion and Summary

We summarize our findings by the addition of two more datasets to the phase diagram shown in Fig. 6.  The first are large, open circles that represent the temperature at which the zone-center, soft mode begins to recover, i.e. the temperature at which the square of the zone-center, soft mode energy begins to vary linearly with temperature as in a conventional ferroelectric material.   The second are the closed circles that represent the values of $T_C$ versus $x$ measured using XRD.   We see that for low PT concentrations the phonon recovery occurs at temperatures well above the C→T phase boundary.   These two temperature scales converge with increasing PT content until they meet near the MPB, becoming a single temperature scale at larger PT concentrations, below which a long-range, ferroelectric phase forms.   This intriguing behavior is consistent with a recently proposed picture of relaxors wherein the structural properties are defined by two temperature scales, which is based on a random-field model in direct analogy to model magnets [45].   Our neutron and x-ray data on PMN-0.32PT show that these two temperature scales merge gradually with PT concentration, and thus point to a dynamical definition of the MPB.   This result is mimicked by the neutron diffuse scattering study of PMN-$x$PT in Ref. [35], which reports that the diffuse scattering is onset at the same temperature for concentrations below the MPB.   The diffuse scattering in that study shows the presence of two temperature scales that merge into one for PT concentrations near the MPB boundary.   For concentrations above the MPB boundary, the temperature dependence of the diffuse component was interpreted in terms of critical scattering, which is expected near a well-defined, continuous phase transition.   The softening of the TO mode there appears to be associated with the formation of polar nanoregions below the MPB and a long-range ferroelectric phase for concentrations above.   The combined results on the phonons and the diffuse scattering show that both are important, and suggest that they are related to the high piezoelectric constants in these materials.

There is no clear effect of the monoclinic transition on the zone-center, soft mode temperature dependence.   This result, combined with those on PMN and PMN-0.60PT, implies a common dynamic response across the PMN-$x$PT phase diagram. Of further interest is the fact that the low-frequency dynamics of PMN, PMN-0.32PT, and PMN-0.60PT are quite different from those of PbTiO$_3$ [25].   PbTiO$_3$ does not show the same degree of damping observed in PMN, PMN-0.32PT, or PMN-0.60PT, and the zone-center, soft mode softens to a much lower energy and recovers at the first order C→T transition. Phonon damping below $T_{Burns}$ has previously been interpreted in terms of a condensation of PNR, which impede the propagation of long-wavelength polar modes [19, 20, 27]. Our results on PMN-0.32PT, combined with those from PMN and PMN-0.60PT, show that this strong damping persists in both relaxor and conventional ferroelectric phases and thus is not a feature unique to relaxors. Instead, the anomalous TO damping seems to correlate directly with the disorder introduced through the extensive mixing of the $Mg^{2+}$, $Nb^{5+}$, and $Ti^{4+}$ cations in these compounds.

In summary, our investigations on PMN-0.32PT, which lies within the compositional range of the MPB, suggest an elegant picture in which the MPB can be defined dynamically as that PT concentration where the zone-center, soft mode recovery and the C→T transition coincide. That the zone-center, soft mode behaves, within our experimental error, identically to that in pure PMN suggests that the low-frequency dynamics in PMN-xPT are universal below the MPB. At higher PT compositions the soft mode behavior is still similar, but scales with $T_C$. In addition, we have found that the boundary between the C and T phases is easily shifted by the application of a small external electric field along [001]. Only in the compositional space inside the MPB, where the transformational sequence C→T→$M_C$ occurs on cooling, is the paraelectric→ferroelectric transition notably shifted higher in temperature with increasing $E_{[001]}$. These results indicate the presence of an incipient structural instability within an apparently cubic phase inside the MPB. The abnormally broad TO mode, which is broader than that in either PMN or PMN-0.60PT, is also suggestive of phase instability. Our findings thus establish the existence of dynamical characteristics that are unique to MPB ferroelectrics.

**Acknowledgments**

We gratefully acknowledge financial support from the Office of Naval Research under Grant N00014-06-1-0204, MURI N00014-06-1-0530, and the Natural Sciences and Engineering Research Council (NSERC) of Canada. We would also like thank Prof. H.

**Figure Captions**

Figure 1 Constant-*Q* scans measured at (2, 2, 0.15) and (2, 2, 0.2) for PMN-0.32PT under $E_{[001]}$=1kV/cm at 450K, 390K, and 100K. The three temperatures correspond to the cubic (C), tetragonal (T), and monoclinic $M_C$ phases, respectively, as identified in the figure 1.

Figure 2 Temperature dependence of the zone-center TO phonon $(\hbar\omega_0)^2$ mode for (a) PMN (zero field); (b) PMN-0.32PT ($E_{[001]}$=1kV/cm); and (c) PMN-0.60PT (zero field). The results for PMN and PMN-0.60PT were measured in the (200) zone, while that for PMN-0.32PT was obtained at (220). The data for PMN and PMN-0.60PT were taken from Stock *et al.* (Refs. [23] and [24]). The closed symbols represent values obtained by extrapolation from nonzero *q* to the zone-center. The open circles were obtained from a direct constant-*Q* scan measured at the zone center.

Figure 3 (a) Dispersion of the TA and TO modes at (220) for PMN-0.32PT at 600K (E=0kV/cm) and 300K ($E_{[001]}$=1kV/cm). The dotted line (or "waterfall" feature [28]) in the 600K data represents the fact that the TO mode becomes heavily damped making it difficult to observe in constant-*Q* scans. (b) Dependence of the TO mode damping on wave vector *q* at 600 K (red) and 300 K (blue), where the zone-center phonon energy was determined through the extrapolation of $(\hbar\omega_0)^2$ as a linear function of $q^2$ to q=0.

Figure 4 Temperature dependence of (a) the TA and TO phonon energy; (b) the TA and TO phonon linewidth Γ. These data were all taken at (2,2,0.15) for PMN-0.32PT under $E_{[001]}$=1kV/cm. The dashed lines through the TO energy and TO phonon linewidths were used to show three changes indication of three phase regions on cooling.

Figure 5 Temperature dependence of (a) the inverse dielectric constant ($1/\varepsilon_r$) at different frequencies under $E_{[001]}$=1kV/cm; (b) the lattice parameters under $E_{[001]}$=1kV/cm; and (c) the *c*-axis lattice spacing measured around the (002) under different electric fields for PMN-0.32PT with E//[001].

Figure 6 Phase diagram of single crystal PMN-$x$PT. The dashed line indicating the transition to the cubic phase was adapted from the phase diagram of PMN-xPT ceramics reported by Noheda *et al.* (Ref. 13), which was the average of the two temperatures reported by Noblanc *et al.*(Ref. 46) from dielectric measurements. The morphtropic phase boundary (MPB) represented by dashed lines (x=0.30 and 0.37) was adapted from Ref. 13. The solid circles denote the structural phase transition temperatures $T_C$ between the cubic and ferroelectric phases after cooling in $E_{[001]}$=1kV/cm. These data were mainly obtained from references [Ref. 13, 18]. The small open circles represent the temperatures at which an abnormal thermal expansion is first observed, which is defined as when the lattice constant begins to increase nonlinearly on cooling in $E_{[001]}$=1kV/cm. The large open circles (with error bars) represent the temperature at which zone-center TO phonon begins to narrow, identified by a minimum in $(\hbar\omega_0)^2$ with temperature. The black thick line through these data denotes the temperatures at which the damping of the soft mode is greatest.

PMN-0.32PT, $E_{[001]}$=1kV/cm, FC

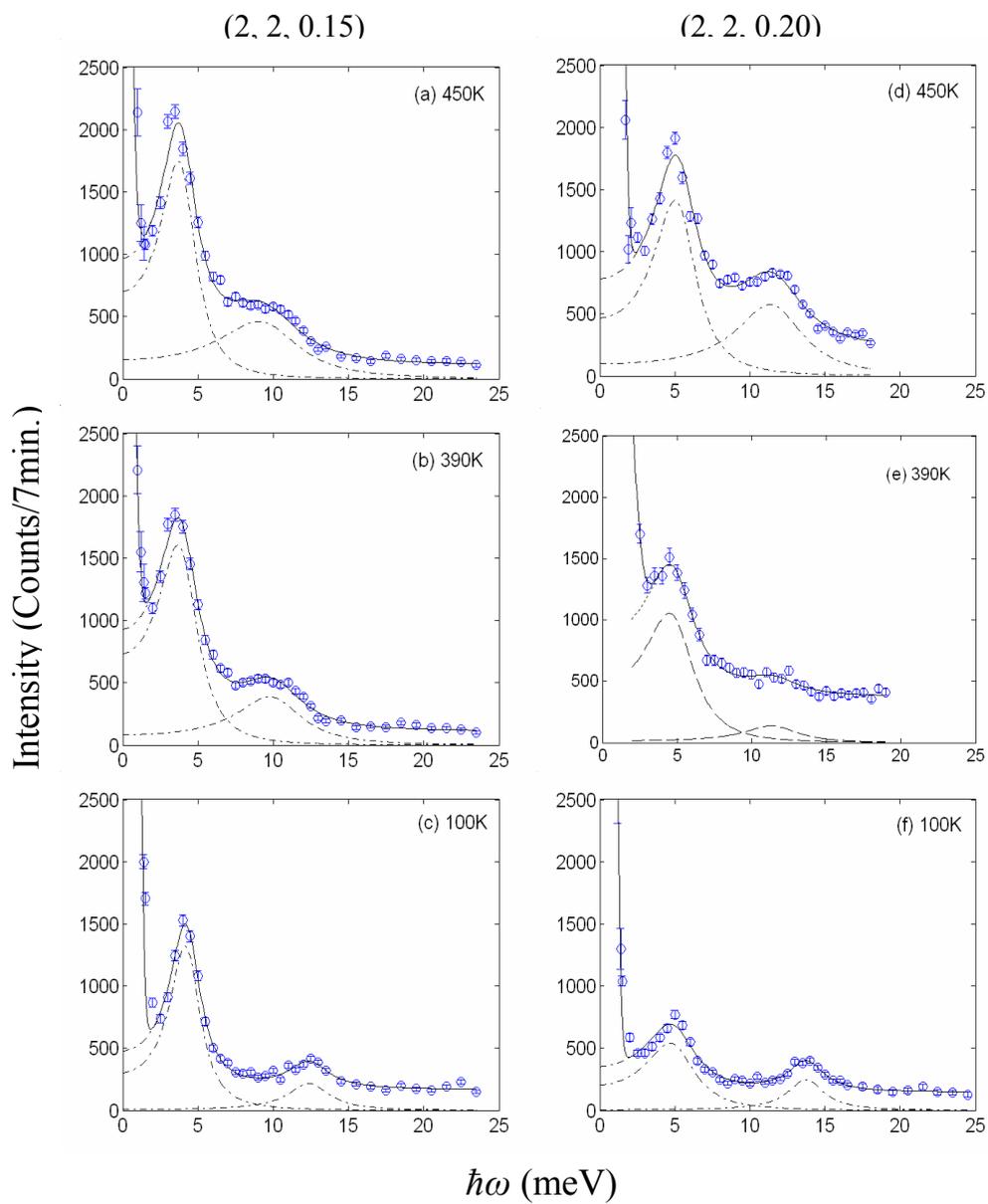

Figure 1

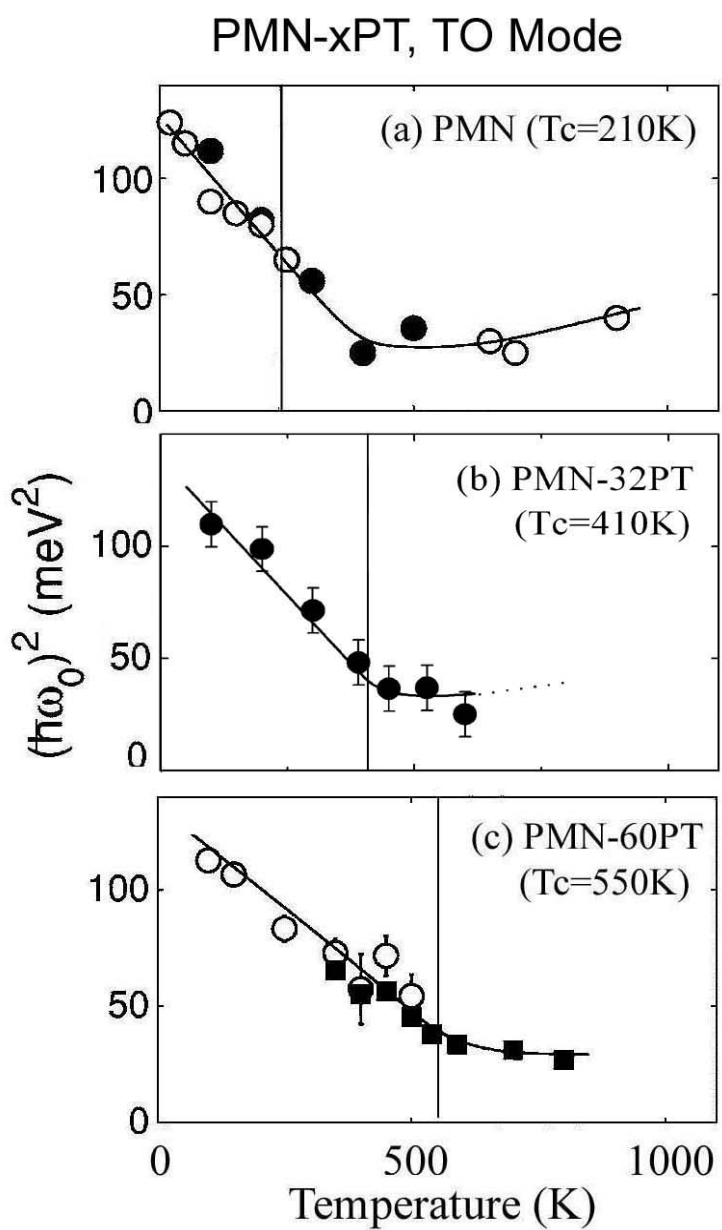

Figure 2

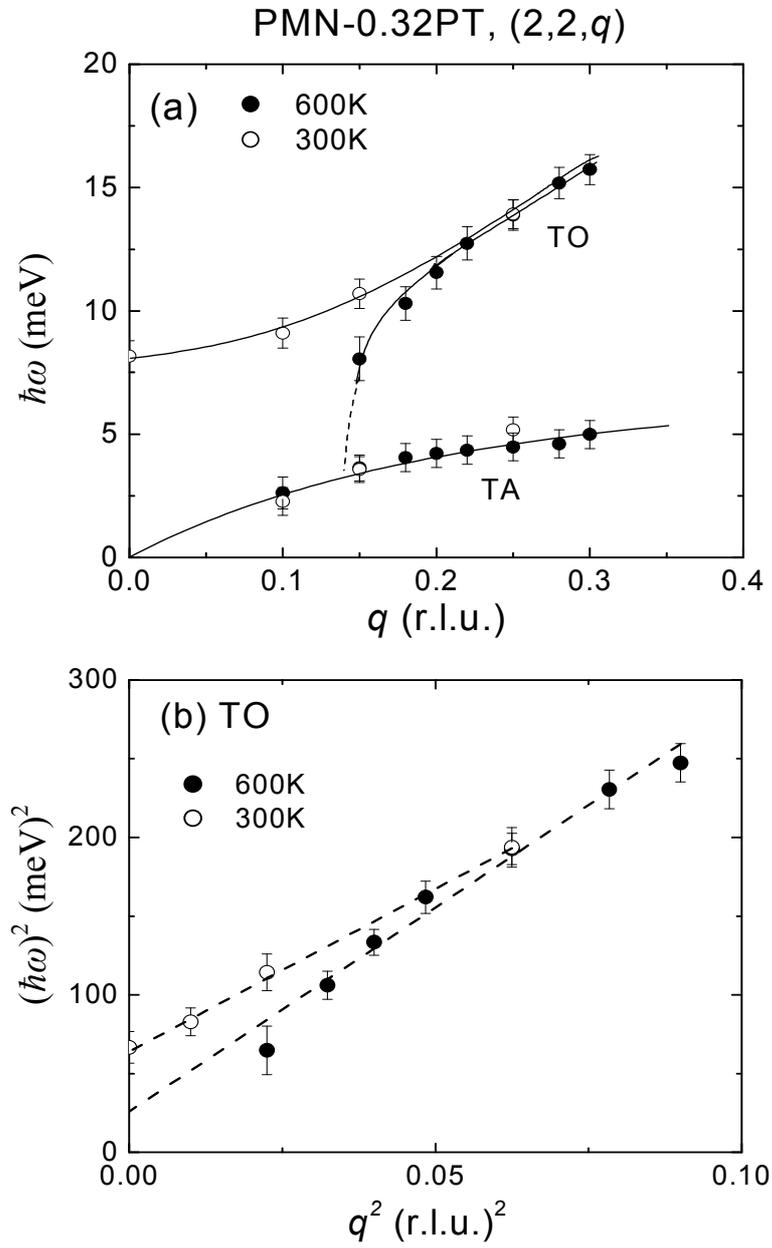

Figure 3

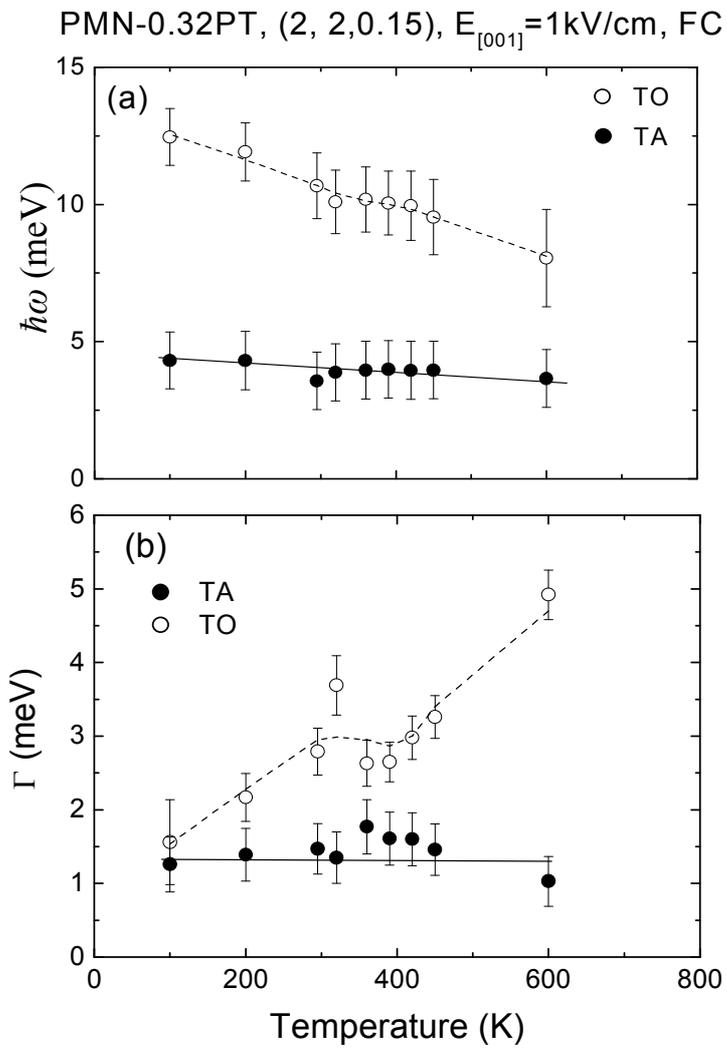

Figure 4

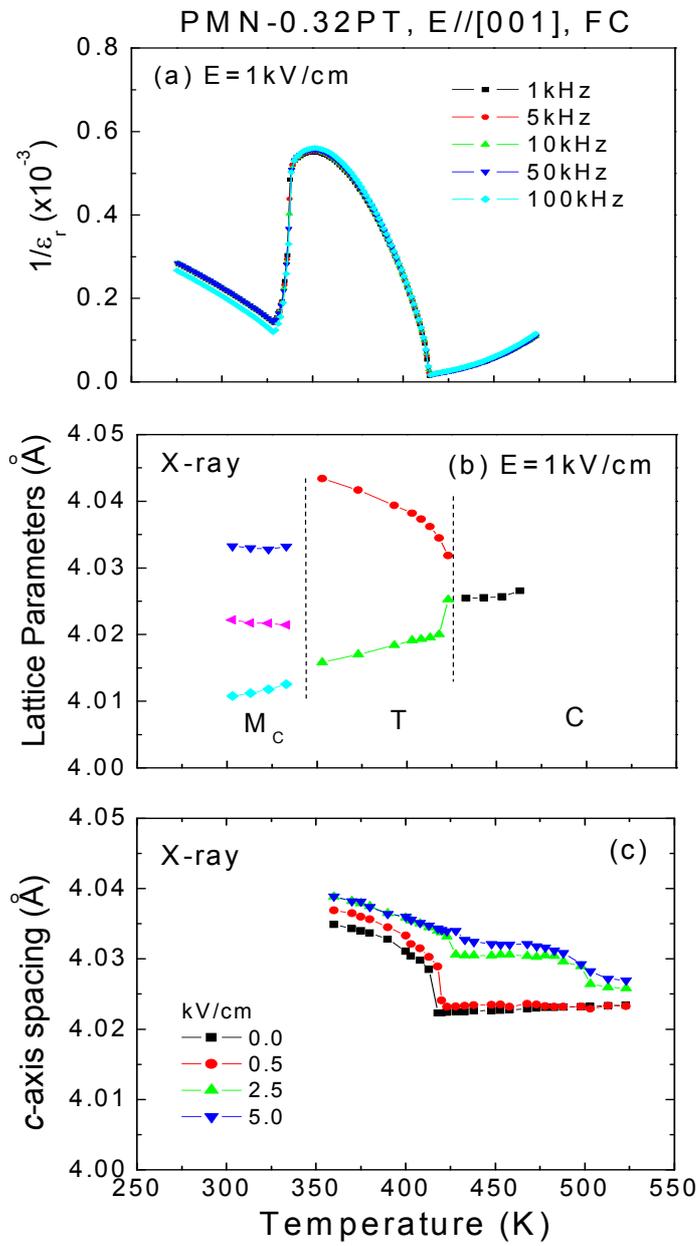

Figure 5

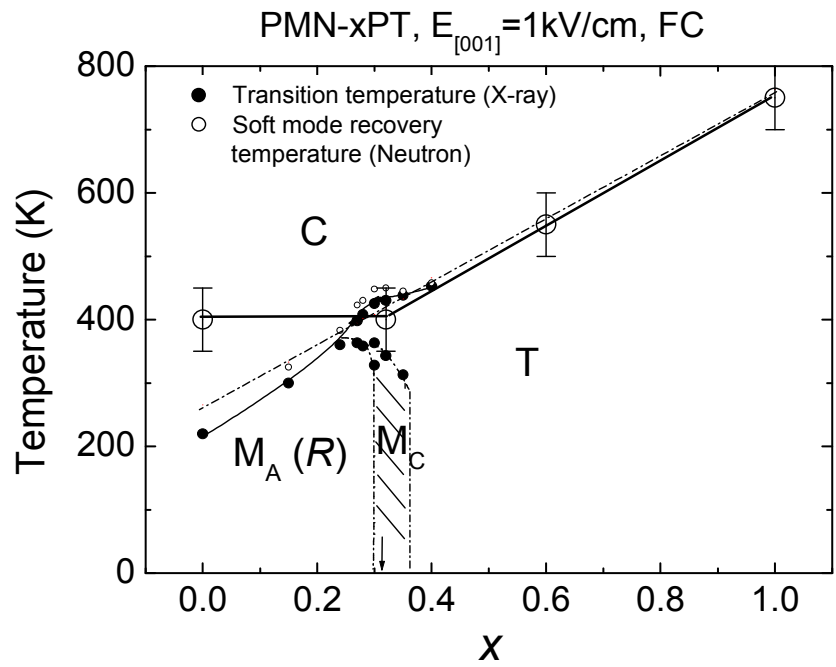

Figure 6